# On the Origin of Below-Bandgap Turn-On Voltage in LEDs and High $V_{OC}$ in Solar Cells comprising Colloidal Quantum Dots with Engineered Density of States

Santanu Pradhan[1], Mariona Dalmases[1] and Gerasimos Konstantatos[1, 2]*

[1]ICFO-Institut de Ciencies Fotoniques, The Barcelona Institute of Science and Technology, 08860 Castelldefels (Barcelona), Spain

[2]ICREA—Institució Catalana de Recerca i Estudis Avançats, Passeig Lluís Companys 23, 08010 Barcelona, Spain

* gerasimos.konstantatos@icfo.es

**ABSTRACT:** The turn-on voltage of a light emitting diode (LED) is an important parameter as it determines the power consumption of the LED and influences the effective power conversion efficiency (PCE). LEDs based on nanoscale engineering of the blended PbS colloidal quantum dots have recently shown record performance in the infrared region. One of the most intriguing results for these blended devices is the substantially lower-than-bandgap turn-on voltage and the achievement of open circuit voltage ($V_{OC}$) - an important figure of merit for photovoltaic (PV) devices – approaching the radiative limit. In this work, we shed new insights on the origin of these phenomena. We posit that the reduction of the effective density of states (DOS) in the conduction and valance band of the emitter QDs in the blended structures modify the chemical potential – that determines the turn-on voltage for LEDs and $V_{OC}$ for PV devices. The reduction of DOS results in an increase of the chemical potential at a particular applied voltage leading to higher photon flux and lower turn-on voltage. Moreover, the change of quasi-Fermi levels due the DOS reduction results in an increase of the $V_{OC}$ in a solar cell device.

Colloidal quantum dots (CQD), in particular lead chalcogenide QD based optoelectronic devices have demonstrated significant potential in optoelectronic applications even competing in performance with current costly epitaxial technologies[1-4]. Solution processed lead sulphide (PbS) CQDs are of particular interest in view of their facile bandgap tunability from the near infrared (NIR) to the short-wave infrared (SWIR)[5]. PbS CQD-based optoelectronic devices such as photodetectors[6], photovoltaics[7] and light emitting diodes[8] (LED) have emerged as a unique low cost alternative to III-V semiconductors being at the same time CMOS compatible. Low cost CMOS compatible infrared Light emitting diodes (LEDs) are highly desired for a large number of application including 3D imaging, on-chip spectrometry etc[9]. To this end, efforts on improving the performance of CQD Infrared LEDs have been reported based on core-shell structures[10], CQD in polymer matrix[11], QD in perovskite matrix[12]. Recently, we have reported another approach based on nanoscale engineering of CQD solids comprising mixed CQD blends that have led to a record PLQE (Photoluminescence quantum efficiency) of >70% in solid state CQD films and peak external quantum efficiency (EQE) of the LED device circa 8% [13]. One astonishing fact of these blend devices has been the observation of a turn-on voltage lower than the bandgap of the emitting CQDs which as a result led to record power conversion efficiency (PCE) of more than 9% - PCE for a LED device is defined as the ratio of the output optical power over the injected electrical power.

Below bandgap turn-on voltage has been traditionally observed in inorganic single crystalline LEDs comprising III-V semiconductors. This has been enabled by carrier diffusion from the electrodes at applied voltage lower than the emission bandgap that radiatively recombine in view of the very low defect density and very high PLQE of these semiconductors. Under these conditions the active material supplies the excess energy through thermal energy to the injected charges followed by device cooling[14-16]. A number of LED devices with different active materials such as polymers[17] or QDs[18,19] have also reported lower-than-bandgap turn-on voltage in which the phenomenon has been hypothesized to be due to Auger assisted charge injection that supplies the extra energy to the injected charges for radiative recombination at the band-edge, yet without conclusive evidence for such hypothesis. In both of the afore-mentioned cases the turn-on voltage of these LEDs has been reported to be lower within 10-15% of the bandgap (1.7 V for 2 eV bandgap).

In our case, the PbS emitter QDs in PbS QD matrix or PbS QD/ZnO NC binary matrix showed lower turn-on voltage compared to the LED devices that comprise only emitter QDs irrespective of the PLQE of the active materials[13]. The reported turn-on voltage in this case is well below prior reports at ~30% of the emission bandgap energy. These findings suggest that another mechanism is at play in this kind of CQD LEDs. We have hypothesized that the reduction of the electronic density of states (DOS) of the emitter QDs due to mixing them in the QD matrix affects the chemical potential in the active emitting sites and thereby enables the radiative transition at an applied forward bias lower than the bandgap. In this report, we have formalized the hypothesis of electronic DOS reduction in blended devices and their influence on the radiation through modifying the chemical potential in the nanocomposite active layer. Then we demonstrate experimentally the effect of the change of electronic DOS of the emitter CQDs on the turn-on voltage in mixed CQD based LED devices. Last but not least we make a connection and correlate the DOS reduction as the common origin of the low turn on voltage in LEDs and the very high $V_{OC}$ recorded in solar cell devices comprising the same mixed QD active layers.

In an LED, electrons and holes are injected from the external circuit to the active semiconducting material to generate light via radiative recombination. The number of injected carriers in the conduction and valance band determine the effective chemical potential in the device. The injected electrons and holes are distributed through the entire conduction and valance band of the active layer where the radiative band-edge recombination results in light emission as shown in Fig. 1(a). In blended devices, emitting low bandgap PbS QD are loaded into a larger bandgap PbS QD based matrix. The emitting QDs form a type-I heterojunction with matrix QDs and the active layer can be assumed as the series of type-I heterojunctions between emitters and the matrix as shown in Fig. 1(b). Injected electrons and holes are transported through the matrix to reach the emitting sites and recombine radiatively. Thus the effective DOS for the emitter only and the emitter in QD matrix based devices are different and depends on the amount of QD loading in the matrix. This reduced effective DOS due to emitter QD loading in the matrix can influence the quasi-fermi level of the active QDs as $n = N_c \exp(\frac{E_{fn}-E_C}{kT})$ , where $N_C$ is the conduction band DOS, $n$ is the number of electrons in the conduction band, $E_{fn}$ is the electron quasi-Fermi level and $E_C$ is the conduction band-edge. To visualize the effect of emitter QD mixing in the QD matrix on the apparent change in quasi-Fermi level *($E_C$-$E_{fn}$),* we have performed SCAPS simulations. For the mixed devices, we have considered it to be as a series of heterojunctions between the emitter QDs and the matrix. The ratio of the matrix and the emitter QD layers were varied according to the QD loading fraction in the matrix (Supporting information S2). Figure 1(c) shows the variation of *($E_C$-$E_{fn}$)* as a function of applied bias and the QD loading fraction (effective DOS reduction fraction). The *($E_C$-$E_{fn}$)* varies with applied bias as well as the DOS reduction. A smaller value of *($E_C$-$E_{fn}$)* indicates the presence of a larger density of injected carriers in the band-edge of the emitting QDs. Simulations show that lowering the fraction of band-edge DOS, leads to lower values of *($E_C$-$E_{fn}$)* compared to the emitter only device. This indicates that with a low fraction QD loading, the quasi-Fermi level of the emitter QD can be much closer to band-edge, even with lower applied bias. On the other hand, the photon flux (Φ) of an LED can be expressed according to Planck's radiation formula as[20],

$$\Phi = \int_0^{\infty} \frac{2E^2}{c^2h^3} \frac{a(E)}{\left\{\exp\left(\frac{E-\mu}{kT}\right)-1\right\}} dE \tag{1}$$

Where $a(E)$ is the absorptivity/emissivity of the active material, $\mu$ is the chemical potential of the radiation at the applied voltage bias, $c$ is speed of light, $h$ is Planck's constant and $k$ is Boltzmann constant. The chemical potential ($\mu$) of an active semiconductor material can be expressed as the sum of electron ($\mu_e$) and hole ($\mu_h$) chemical potentials and can be written as the difference between the quasi-Fermi levels as[21],

$$\mu = \mu_e + \mu_h = E_{fn} - E_{fp} \tag{2}$$

Thus, the change of electronic DOS in the band-edge can modify the quasi-Fermi levels and as a consequence it can modify the chemical potential of the active layer. Equation (1) shows the overlap integral of $a(E)$ and the radiation curve which is governed by $\mu$, determines the photon flux. Figure 1(d) shows the evolution of the overlap integral with the change of $\mu$. The gradual increase of $\mu$ towards the bandgap value results in increasing the overlap integral and hence the photon flux. The effect of $\mu$ and $a(E)$ on the emitted photon-flux is shown in Fig. 1(e). It depicts a more prominent influence of $\mu$ on photon-flux compared to $a(E)$ which varies due to the different blending ratios. Thus, the DOS reduction can enhance the photon-flux from the LED devices through the change of $\mu$ irrespective of the emissivity of the materials. With DOS reduction, higher $\mu$ can be achieved for blended devices compared to the emitter only device at a particular applied voltage and thus can show a lower turn-on voltage.

In order to experimentally verify our model we fabricated LED devices. Figure 2(a) shows the energy levels of the different layers in a typical mixed CQD LED device. Electrons are injected from indium tin oxide (ITO) electrode and holes are injected from Au electrode. ZnO is the electron injecting and hole-blocking layer. The blended QDs treated with $ZnI_2$ and 3-Mercaptopropionic acid (MPA) based mixed ligand as we have reported earlier[22], are the active material for radiative recombination. 1,2-Ethanedithiol (EDT) treated PbS QD of same size as the matrix was used as the hole transporting and electron blocking layer similar to previously reported structures for PbS QD based photovoltaics[23] (details are given in the Supporting information S1). The optimized performance of the blend configuration and emitter only devices are shown in supporting information S3. The optimized loading of 7.5% emitter QDs in matrix QDs showed a peak EQE (external quantum efficiency i.e. the ratio of emitted photons to the injected charge carriers) of 5% compared to the EQE of 0.55% for the emitter only based device. Interestingly, the turn-on voltage of the QD blend devices was measured around 0.6 V whereas that of the emitter only device was around 0.9 V, which is close to the bandgap equivalent of the emitting QDs (~0.87 eV). To observe the effect of electronic DOS on the turn-on voltage of the devices, we have prepared a series of binary blend devices with varying DOS (emitter QD loading fraction). The turn on voltage of the LED devices reduces progressively upon increasing reduction of the emitter QD electronic DOS in the active layer as shown in Fig. 2(c). Furthermore, we wanted to confirm that this lower than bandgap turn-on voltage is mainly due to DOS reduction and not the result of variation in the PLQE. To do so, we have prepared devices with 0.002% MPA_$ZnI_2$ based mixed ligand treatment, which showed 5-20 times lower PLQE compared to the optimized blends depending on the blending ratio and is herein called low PLQE film. Figure 2(b) shows the PLQE dependence on the emitter QD fraction in the blend for the high PLQE and the low PLQE films. PLQE for both cases increases with reducing the emitter QD fraction due to the probability reduction of charge capture in defective dots[13]. The low PLQE film exhibits a dramatic increase in PLQE when the fraction of emitter QDs drops below 60%. However the corresponding reduction of the turn-on voltage (Fig. 2(c)) onsets at a loading fraction of 80%, i.e. in the low PLQE regime. Moreover, despite the two films having a different PLQE value at the lowest studied fraction (10% for the low PLQE film and 53% of the high PLQE film) they exhibit an almost identical value of turn-on voltage. By comparing the two films at equivalent PLQE values (which occurs at 20% emitter QD loading for the low PLQE film and 60% emitter QD loading for the high

PLQE film) we note a significant difference in their corresponding turn-on voltage values (0.65 V and 0.77 V respectively). This may discard the PLQE as the main contributing factor to the low-turn on voltage effect and further corroborates our hypothesis of the reduced DOS and the modification of the chemical potential as the underlying mechanism for this.

To test this model we have experimentally calculated the value of $\mu$ from radiation plot and observed the effect of DOS modification on $\mu$. The $\mu$ of radiative recombination in an LED is given as[21],

$$\mu = \hbar\omega + kTln(\frac{4\pi^2 c^2 \hbar^3 I(\hbar\omega)}{a(\hbar\omega)(\hbar\omega)^3}) \quad (3)$$

Where $\hbar\omega$ is the average energy of the emitting photons, $a(\hbar\omega)$ is the absorptivity/emissivity, and $I(\hbar\omega)$ is the intensity of the radiation given by,

$$I(\hbar\omega) = \frac{i_e(\hbar\omega)\eta}{qAw} \quad (4)$$

Where $i_e$ the injected current, $\eta$ is the EQE at that applied voltage, $A$ is the emitting device area, and $w$ is the FWHM of the emission spectra. The $\mu$ of the radiation at different QD DOS fraction can be calculated using equation (3) and (4) (the detailed calculation is shown in supporting information S5). On the other hand, the change of chemical potential ($\Delta\mu$) due to the change of conduction and valance band DOS can be written as,

$$\Delta\mu = kTln(\frac{N_{C1}N_{V1}}{N_C N_V}) \quad (5)$$

Where the conduction band DOS changes from $N_C$ to $N_{C1}$ and valence band DOS changes from $N_V$ to $N_{V1}$ due to QD blending. So, the value of $\Delta\mu$ can be predicted from the change of DOS in a semiconducting system considering there is no significant change in charge injection. Figure 2(d) shows the predicted and experimentally extracted value of $\Delta\mu$ as a function of DOS reduction fraction. The similarity between experimental and predicted value further confirms the effect of DOS reduction on $\mu$ enhancement and as a consequence the increment of photon flux at lower voltage bias which gives rise to the below bandgap turn-on voltage.

Finally, we wanted to correlate the low turn on voltage with the high $V_{OC}$ previously reported in these devices[13] and have therefore prepared photovoltaic (PV) devices based on the blend structures. The PV device structures follow a similar layer formation of LED devices yet with a much thicker active layer of approximately 200 nm and thinner blocking layer of approximately 35 nm to facilitate efficient photon absorption and charge collection. The open circuit voltage ($V_{OC}$) of the devices are directly proportional to the difference of quasi-Fermi levels. Figure 3(a) shows the change of $V_{OC}$ as a function of emitter QD loading in the QD matrix. The emitter only QD based device showed a $V_{OC}$ of 0.42 V. The reduced emitter DOS fraction of 80%, 60%, 40%, and 20% resulted in the $V_{OC}$ of 0.48 V, 0.52 V, 0.56 V, and 0.59 V respectively. Figure 3(b) shows the external quantum efficiency (EQE) of the PV devices. The excitonic peak intensity around 1320 nm (corresponding to the emitter QD absorption) changes according to the fraction of emitter QD loading in the blend device confirming the emitter QD DOS reduction in these blends. A similar type of $V_{OC}$ increment in the PbS QD blends was shown by some previous reports and ascribed to the modification of quasi-Fermi level due to DOS reduction and variation in charge transport[13,24]. To quantify these effects on the $V_{OC}$, we have performed wavelength dependent $V_{OC}$ measurements as shown in Fig. 3(c). We have selected the incident-light wavelengths of 640 nm and 1320 nm (same intensity) where 640 nm can excite both the matrix and the emitter QDs whereas 1320 nm light can only excite the emitter QDs in the matrix. The $V_{OC}$ of the blended devices, obtained upon 1320 nm light excitation, showed gradual increase with reduced DOS. There are two major mechanisms that contribute this change in $V_{OC}$. One of the mechanisms is the reduction of DOS and the other one is the charge carrier reduction due to mixing. The contribution of the reduction of DOS on the change of quasi-Fermi levels which governs the $V_{OC}$, can be estimated from equation (5) and shown in Fig. 3(c). The rest of the contribution is coming from the charge carrier reduction due to mixing. When the devices were excited with 640 nm laser, the $V_{OC}$ increased further due to the contribution of photo-generated carriers from the matrix QDs. SCAPS simulations (details are in supporting information S6) of the PV devices further support the change of $V_{OC}$ due to QD mixing. We have considered the mixed devices as a series of type-I heterojunctions between the emitter QDs and the matrix QDs for the simulation and the ratio of the thickness between the emitter and QD matrix varied according to the emitter QD loading fraction in the matrix. The simulated $V_{OC}$ as a function of band-edge DOS fraction and the active layer mobility is shown in Fig. 3(d). It shows that the variation of mobility (due to the mixing of QDs) does not influence much change in $V_{OC}$, whereas the DOS reduction inflicts a distinct change in device $V_{OC}$. The agreement of the SCAPS modeling with the experimental results further corroborates the role of DOS reduction on the enhancement the $V_{OC}$ of the PV devices through the modification of quasi-Fermi levels.

We have successfully showed the electronic DOS modification with nanoscale engineering of the solution processed quantum dots can be used to tune the chemical potential in semiconducting systems which can influence the turn-on voltage of the LED devices and the $V_{OC}$ of the PV devices. We have achieved nearly 0.3V lower than bandgap turn-on voltage with 0.075 DOS fraction and $V_{OC}$ of 0.59V from 0.42V upon a fivefold reduction of the DOS in PV device.

**Supporting information:**

The supporting information contains:

*Experimental methods; SCAPS simulation to visualize quasi-Fermi level change as a function of DOS reduction; LED device performance; Radiance plot as a function of DOS; Chemical potential calculation from radiation curve; SCAPS simulations for PV devices; SCAPS simulation parameters.*

**Acknowledgements**

The authors acknowledge financial support from the European Research Council (ERC) under the European Union's Horizon 2020 research and innovation programme (grant agreement no. 725165), the Spanish Ministry of Economy and Competitiveness (MINECO), and the "Fondo Europeo de Desarrollo Regional" (FEDER) through grant TEC2017-88655-R. The authors also acknowledge financial support from Fundacio Privada Cellex, the program CERCA and from the Spanish Ministry of Economy and Competitiveness, through the "Severo Ochoa" Programme for Centres of Excellence in R&D (SEV-2015-0522).

## Figures

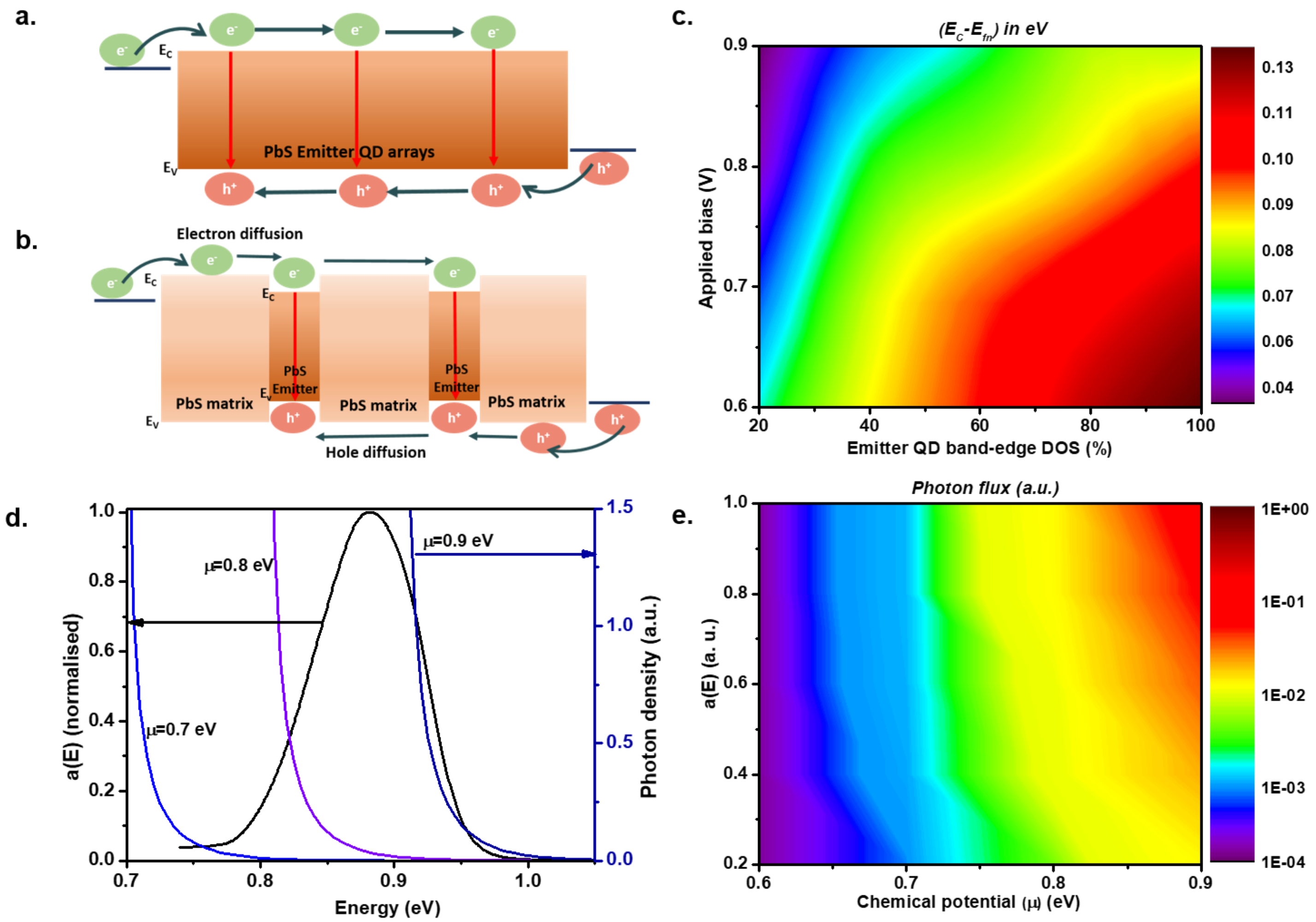

**Figure 1: Effect of DOS reduction chemical potential and photon-flux of LED devices. a. b.** Energy level schematic of DOS reduction due to quantum dot mixing. In emitter QD only device, the electrons and holes are injected to the entire span of conduction and valence band and the band-edge recombination take place throughout the arrays (**a**). On the other hand, the blended bulk heterojunction of matrix and emitter can be considered as the series of type-I heterojunctions between the matrix and emitter QDs. The injected charges diffused through the matrix and the radiative recombination take place only at the emitter QD sites (**b**). Thus, the electronic DOS reduced due to the blending of QDs. **c.** The variation of electron quasi-Fermi level ($E_{fn}$) from the conduction band ($E_C$) as a function of DOS reduction and applied voltage determined by SCAPS simulation. The difference is taken as the average over the entire region of the emitter QD layer. Strong dependence of DOS reduction on ($E_C$-$E_{fn}$) observed. **d.** Overlap of photon density and emissivity (a(E)) to calculate photon flux. Increase of the chemical potential towards band-edge energy enhances the overlap integral. **e.** Emitting photon flux in a standard LED device as a function of chemical potential and a(E). Strong dependence of the photon flux on chemical potential observed.

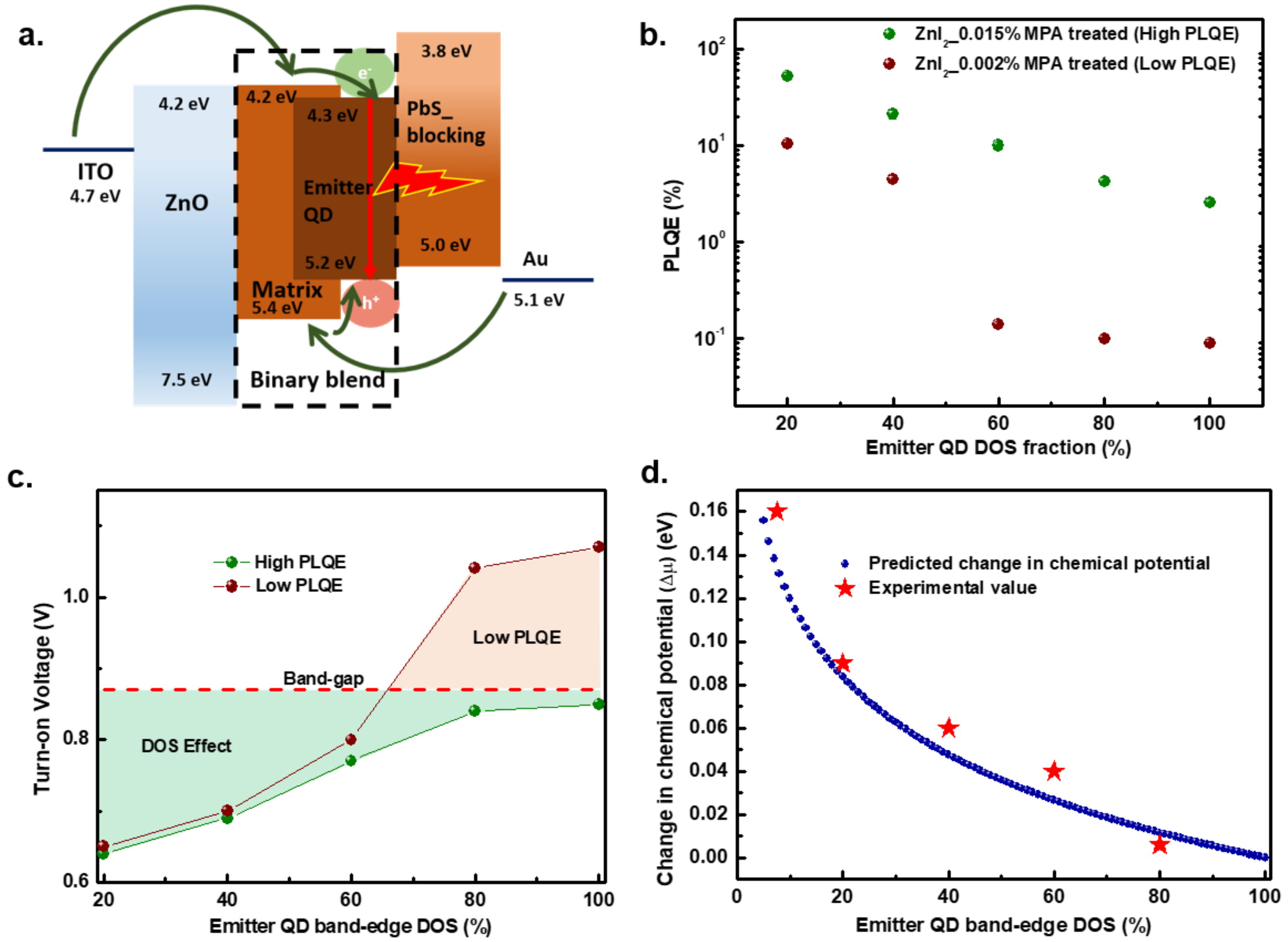

**Figure 2: Effect of DOS reduction on the turn-on voltage of the LED devices: a**. Schematic energy band diagram of blended LED devices. In blended device, the matrix and emitter QDs form a type-I heterojunction throughout the blend. **b.** The variation of PLQE as function of emitter QD loading fraction for two different concentrations of MPA treatment. The low PLQE film ($ZnI_2$_0.002% MPA treated) showed extremely low PLQE value with high emitter loading fraction. **c.** Variation of turn-on voltage as a function of emitter QD band-edge DOS fraction for LEDs based on high and low PLQE materials. Very low PLQE hinders the observation of below-bandgap turn-on voltage (red shaded region). On the other hand, in the higher PLQE region, turn-on voltage showed prominent influence of DOS (green shaded region) **d.** Predicted change of chemical potential as a function of DOS reduction. The red stars show the experimental values extracted from the radiance plot. The experimental values follow the similar trend predicted by DOS reduction.

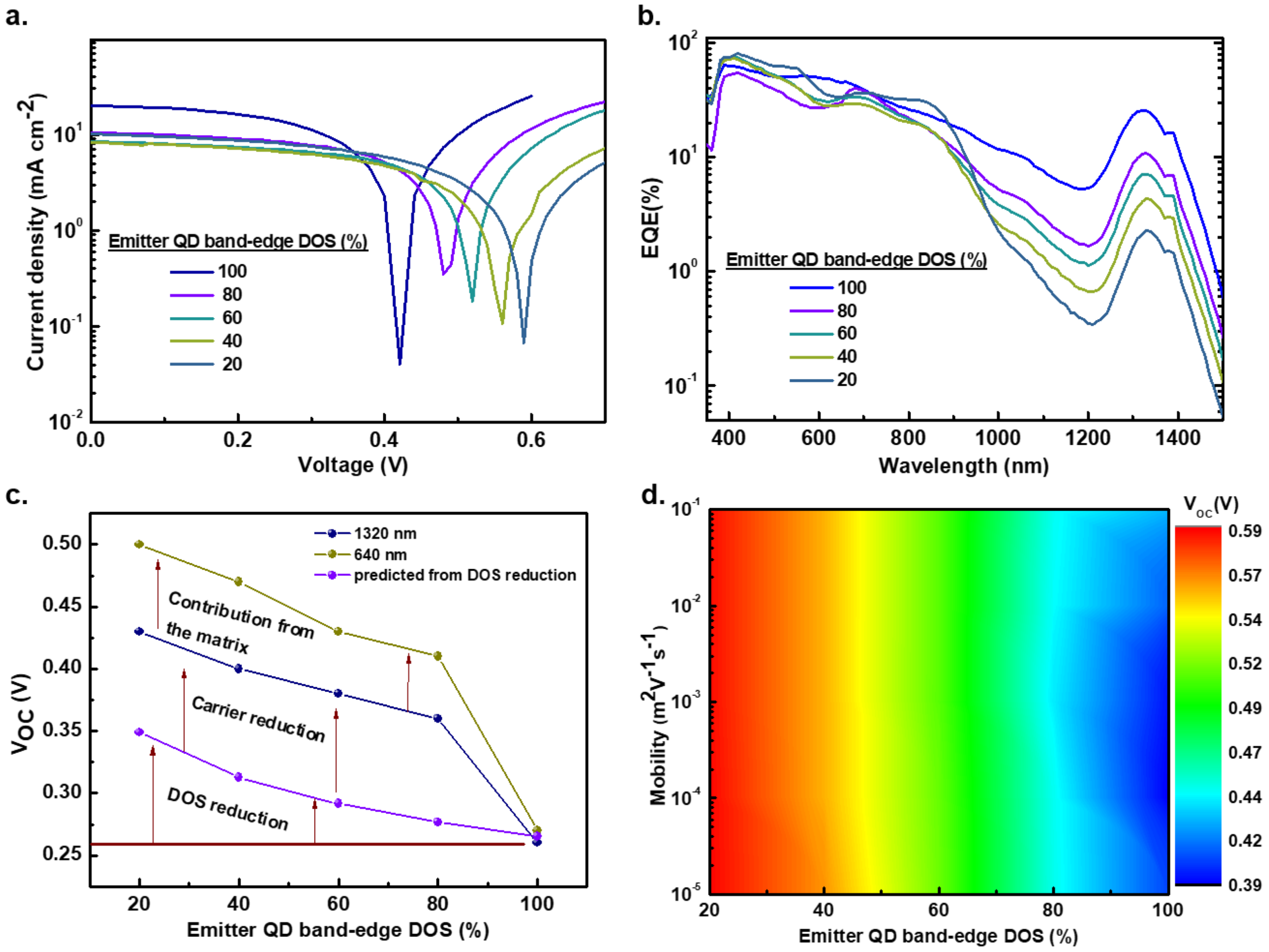

**Figure 3: PV performance as a function of DOS reduction: a.** Photovoltaic performance as a function of DOS variation. Systematic change of $V_{OC}$ is observed with DOS reduction. **b.** EQE of the PV devices. The excitonic peaks around 1320 nm show the loading of emitter QD in the matrix which confirms the electronic DOS reduction of the emitter QDs with different blend mixing. **c.** $V_{OC}$ variation as a function of DOS and wavelength variation. We have considered two different wavelengths of 640 nm (excites both matrix and emitter) and 1300 nm (only excites emitter QDs). The effect of DOS from theoretical calculation shown in violet curve. The contribution from the carrier reduction due to mixing and matrix QD absorption on the $V_{OC}$ is shown. **d.** SCAPS simulated $V_{OC}$ of the devices as a function of DOS reduction and mobility. The clear influence of DOS on $V_{OC}$ is observed.

**TOC:**

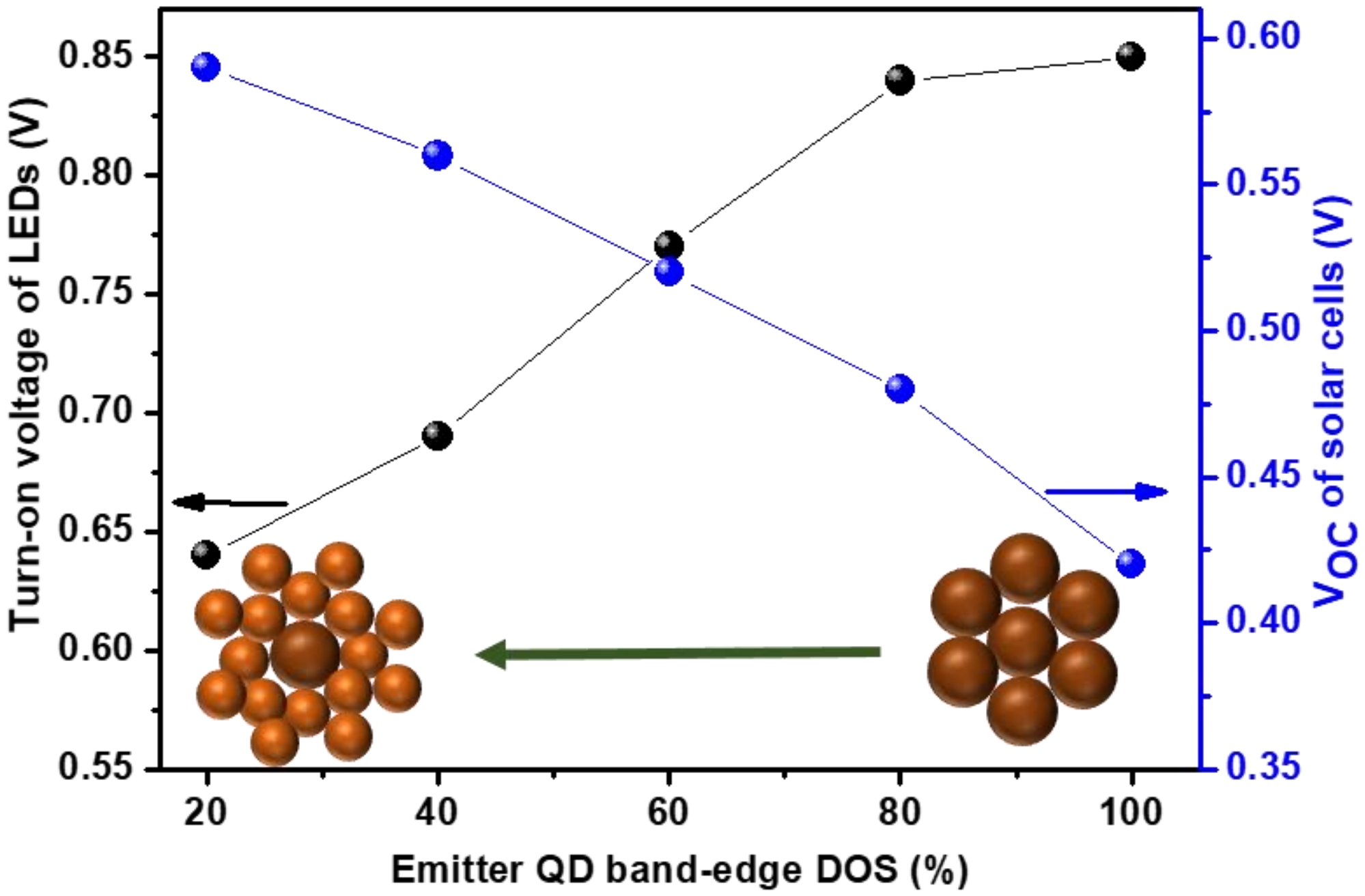